\documentclass[aps,prl,showpacs,amssymb,superscriptaddress,footinbib,twocolumn]{revtex4} 
\usepackage{graphicx}
\usepackage{amsmath}
\usepackage{color}
\usepackage{cancel}
\usepackage{ulem} 
\usepackage{marginnote}
\usepackage{tikz}
\newcommand{\ket}[1]{\ensuremath{\left|{#1}\right\rangle}}

\newcommand\constructosum[3]{%
    \begin{tikzpicture}[baseline=(char.base), inner sep=0, outer sep=0]
        \draw (#1,0) circle (#2); 
        \node (char) at (0,0) {$#3\sum$}; 
    \end{tikzpicture}%
}
\newcommand{\modtwosum}{\mathop{\mathchoice
        {\constructosum{-0.3ex}{0.1}{\displaystyle}}
        {\constructosum{-0.3ex}{0.06}{\textstyle}}
        {\constructosum{-0.2ex}{0.04}{\scriptstyle}}
        {\constructosum{-0.15ex}{0.03}{\scriptscriptstyle}}
    }\displaylimits
}

\begin{document}

\title{Extracting past-future vacuum correlations using circuit QED}

 \author{Carlos Sab\'in}
\affiliation{Instituto de F\'{\i}sica Fundamental, CSIC,
  Serrano 113-B, 28006 Madrid, Spain}
\email{csl@iff.csic.es}

\author{Borja Peropadre}
\affiliation{Instituto de F\'{\i}sica Fundamental, CSIC,
  Serrano 113-B, 28006 Madrid, Spain}
  
\author{Marco del Rey}
\affiliation{Instituto de F\'{\i}sica Fundamental, CSIC,
  Serrano 113-B, 28006 Madrid, Spain}

\author{Eduardo Mart\'{i}n-Mart\'{i}nez}
\affiliation{Instituto de F\'{\i}sica Fundamental, CSIC,
  Serrano 113-B, 28006 Madrid, Spain}
\affiliation{Institute for Quantum Computing, Department of Physics and Astronomy and Department of Applied Mathematics, University of Waterloo, 200 University
Avenue W, Waterloo, Ontario, N2L 3G1, Canada}

\date{\today}

\begin{abstract}
We propose a realistic circuit QED experiment to test the extraction of past-future vacuum entanglement to a pair of superconducting qubits.  The qubit $P$ interacts with the quantum field along an open transmission line for an interval $T_\text{on}$ and then, after a  time-lapse $T_\text{off}$, the  qubit $F$  starts interacting for a time  $T_\text{on}$ in a symmetric fashion.  After that, past-future quantum correlations will have transferred to the qubits, even if the qubits do not coexist at the same time. We show that this experiment can be realized with current technology and discuss its utility as a possible implementation of a quantum memory.
\end{abstract}

\maketitle
\textit{Introduction.--} 
The fact that the vacuum of a quantum field presents quantum entanglement was discovered long ago \cite{Unruh,summerswerner, summerswernerII}, but it was considered a mere formal result until it was addressed from an applied perspective in \cite{reznik}. Since then, this intriguing property has attracted a great deal of attention as a possible new resource for Quantum Information tasks \cite{reznikII, reznikIII, cqedsabin, clichekempf}.

As shown in \cite{reznik}, the entanglement contained in the vacuum of a scalar field can be transferred to a pair of two-level spacelike separated detectors interacting with the field at the same time. Unfortunately, this theoretical result seems to be very difficult to translate into an experiment, even in the context of a trapped-ion simulation \cite{reznikII}. 
Recently, it has also been proven \cite{olsonralphfp} that the vacuum of a massless scalar field  contains quantum correlations \cite{comment} between the future and the past light cones. A theoretical method of extraction by transfer to detectors interacting with the field at different times has also been proposed \cite{olsonralphpfextraction},  but the particular time dependence of the energy gaps seems extremely challenging from the experimental viewpoint. Another ideal proposal  was provided in \cite{ivettedrago} with a setting that seems even more difficult to tackle experimentally.

On the other hand, circuit QED \cite{reviewnature} provides a framework in which the interaction of two-level systems with a quantum field can be naturally considered. The combination of superconducting qubits with transmission lines implement an artificial 1-D matter-radiation interaction, with the advantage of a large experimental accessibility and tunability of the physical parameters. Using these features, fundamental problems in Quantum Field Theory hitherto considered as ideal are now accessible to experiment \cite{wilson}. In particular, the possibility of achieving an ultrastrong coupling regime  \cite{bourassa09,forn-diaz10,niemczyk10}  has already been exploited to propose a feasible experimental test of the extraction of vacuum entanglement to a pair of spacelike separated qubits \cite{cqedsabin}. 

\begin{figure}[t!] 
\includegraphics[width=\columnwidth]{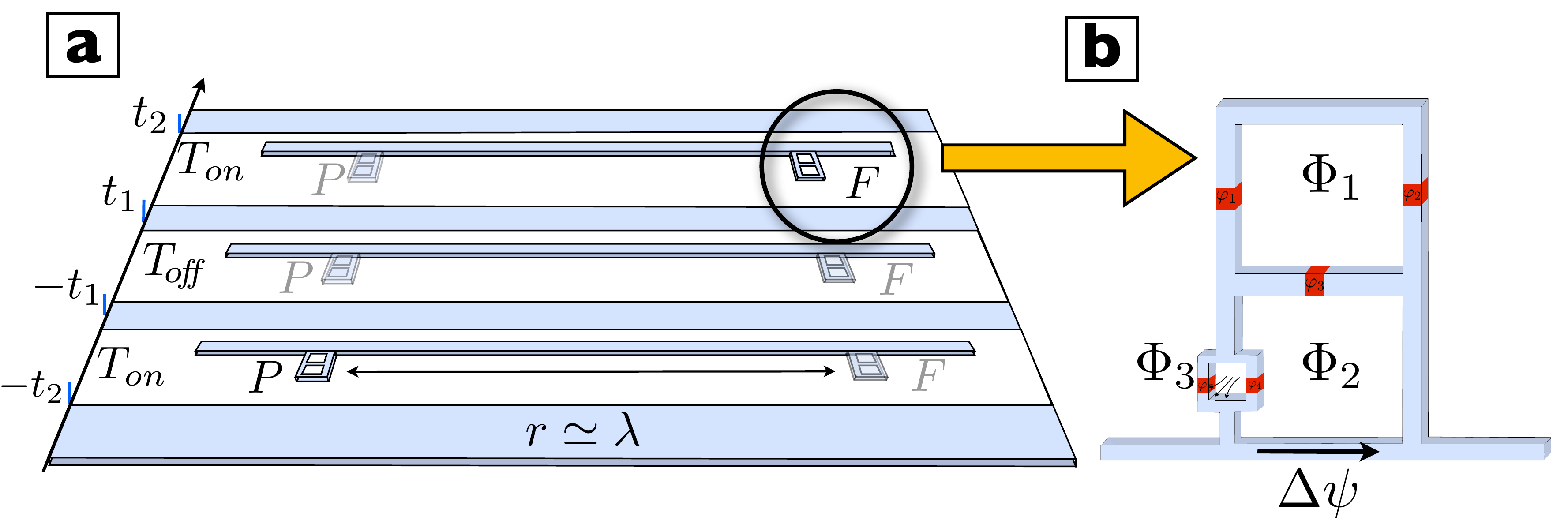}
\caption{Experimental proposal for past-future entanglement
  extraction. a) Time evolution of our protocol: the qubit P interacts
  with the vacuum field ($\Delta\psi$) for a time $T_\text{on}$. After
  a certain time $T_{\text {off}}$ with no interaction, a second qubit
  F interacts with the field getting entangled with the qubit P. b)
  Switchable coupling design: a flux qubit (top ring) is coupled to the field $\Delta\psi$ by ways of two loops. Varying the magnetic
  fluxes $\Phi_2$ and $\Phi_3$ we deactivate the qubit-field coupling.}
\label{fig:imp}
\vspace{-0.5 cm}
\end{figure}

In this work we will take advantage of the aforementioned features of circuit QED in the ultrastrong coupling regime in order to propose a realistic experiment for the extraction of past-future correlations \cite{comment2} contained in the vacuum of a quantum field.  We will consider a set-up consisting of a pair of superconducting qubits $P$ and $F$ with constant energy gaps in a common open transmission line (Fig.~\ref{fig:imp}a). First,  the interaction  of $P$ with the vacuum of the field is on for a time interval $T_\text{on}$ (we call this interval `the past').  Then, $P$ is disconnected from the field during a  time $T_\text{off}$. Finally, the interaction of $F$ is switched on during $T_\text{on}$  (`the future') while keeping $P$ disconnected. After this procedure, we will show that the qubits can end up in a  strongly correlated quantum state, in spite of not having interacted with the field at the same time. We will consider three different spacetime configurations: that the qubits are spacelike or timelike separated, and in the latter case with or without photon exchange allowed. Perhaps the most surprising result is that, even if photon exchange is forbidden,  the qubits can get entangled by a transference of vacuum correlations, as we will show. However, this is not the only interesting aspect of our scheme. If there is a certain probability of photon exchange, some classical correlations between the qubits are obviously expected. But is also remarkable that, due to the peculiarities of our circuit QED setup these correlations are quantum and attain a high degree without the need of a projective measurement of the field. We stress that our proposal is free of idealized requirements such as gaps with unfeasible time dependences. Our switching scheme is fully within reach of current circuit QED technologies, as shown below. 

Our protocol has also an important applied counterpart. As suggested in \cite{olsonralphpfextraction}, the extraction of past-future quantum correlations enables its use as a quantum channel for quantum teleportation ``in time''. We will show how this opens the door  to a novel kind of quantum memory in which the information of the quantum state of some ancillary qubit $P'$ is codified in the  field during $T_\text{off}$ and then recovered in $F$ using classical information stored in the past - regardless whatever may happen to $P$ after its interaction with the field. 

\textit{Theoretical model.--} 
We focus on a  setup of circuit-QED with two superconducting qubits $P$ and $F$
interacting via a quantum field. The states
$\ket{e}$ and $\ket{g}$ are separated by a constant energy $\hbar\Omega$. The 1D field, $\Delta\psi(x),$ propagates
along an open microwave guide or transmission line that connects them
\begin{eqnarray}
  \Delta\psi(x)=i\,\int_{-\infty}^{\infty}dk\, \sqrt{N\omega_k}\,e^{ikx}a_k +\mathrm{H.c.} \label{a}
\label{field}
\end{eqnarray}
This field has a continuum of Fock operators $[a_k,a^{\dag}_{k'}]=\delta(k-k'),$ and a linear spectrum, $\omega_k =
v|k|$, where $v$ is the propagation velocity of the field. The normalization $N$ and the speed of photons, $v=(cl)^{-1/2},$ depend on the microscopic details such as the capacitance and inductance per unit length, $c$ and $l.$ We will assume qubits that are much smaller than the relevant wavelengths, $\lambda=2\pi\,v/\Omega_J,$ ($J=P,F$) and the fixed distance $r$. Thus the Hamiltonian, $H = H_0 + H_\text{I},$  splits into a free part
 $ H_0 = \frac{1}{2}\hbar(\Omega_P\sigma^z_P + \Omega_F\sigma^z_F) + \int_{-\infty}^{\infty}dk\, \hbar\omega_k  a^{\dagger}_ka_k \label{b}$
and a point-like qubit-field interaction:
\begin{equation}
  H_\text{I} = - \sum_{J=P,F} d_J\,\Delta\psi(x_J)\,\sigma_J^x =H_\text{IP}+H_\text{IF}.\label{c}
\end{equation}
Here $x_J$  are the fixed positions of the atoms, and $d_J\, \sigma^x_J$ comes from a dimensional reduction of the matter- radiation interaction hamiltonian with two-level atoms and the electromagnetic field, analogous - but not fully equivalent - to the Unruh-de Witt model. \cite{DeWitt}. 

We choose the following initial state  $|\Psi(-t_2)\rangle = |eg0\rangle,$ where only $P$ has been excited, in order to analyze the interplay between photon exchange and vacuum correlations effects in the generation of entanglement. According to our past-future scheme (Fig.~\ref{fig:imp}a), the system evolves in the interaction picture into the state
\begin{equation}
  \ket{\Psi(t_2)} = {\cal T}e^{-i \int_{-t_2}^{t_2} \frac{dt'}{\hbar} [\Theta(-t'-t_1) H_\text{IP}^{(t')}+\Theta(t'-t_1) H_\text{IF}^{(t')}]}\ket{eg0},\label{d}
\end{equation}
${\cal T}$ being the time ordering operator. 

We use the formalism of perturbation theory  up to the second order and beyond Rotating Wave Approximation \cite{cqedsabin}  and trace over the field degrees of freedom to obtain the corresponding two-qubit reduced density matrix $\rho_{PF} $ evaluated at $t_2$. The degree of entanglement of this X-state can be characterized with the concurrence, which is given by: $ \mathcal{C}(\rho_{PF})=2 \Big[|X|-\Big({\sum_{k}|A_{1,k}|^2 \sum_{k}|B_{1,k}|^2}\Big)^{1/2} \Big]$,
$X$ standing for the amplitude of photon - real and virtual - exchange and $\sum_{k}|A_{1,k}|^2$,  $\sum_{k}|B_{1,k}|^2$ for the probability of single-photon emission by $P$ and $F$, respectively. These terms can be computed - following similar techniques as in~\cite{cqedsabin} - as a function of four dimensionless parameters, $\xi_\text{on}$, $\xi_\text{off}$, $K_P$ and $K_F$.  The first two, $\xi_\text{on}=vT_\text{on}/r$, $\xi_\text{off}=vT_\text{off}/r$ allow to discriminate the different spacetime regions. The remaining ones are dimensionless coupling strengths for  qubits P and F:  $K_J=4d_J^2N/ (\hbar^2 v)= 2\left(g_J/\Omega_J\right)^2.$ We will restrict to $2 K_J\Omega_Jt_2 \ll 1 $ where our perturbative approach to remain valid.

\begin{figure*}[t] 
\includegraphics[width=\textwidth]{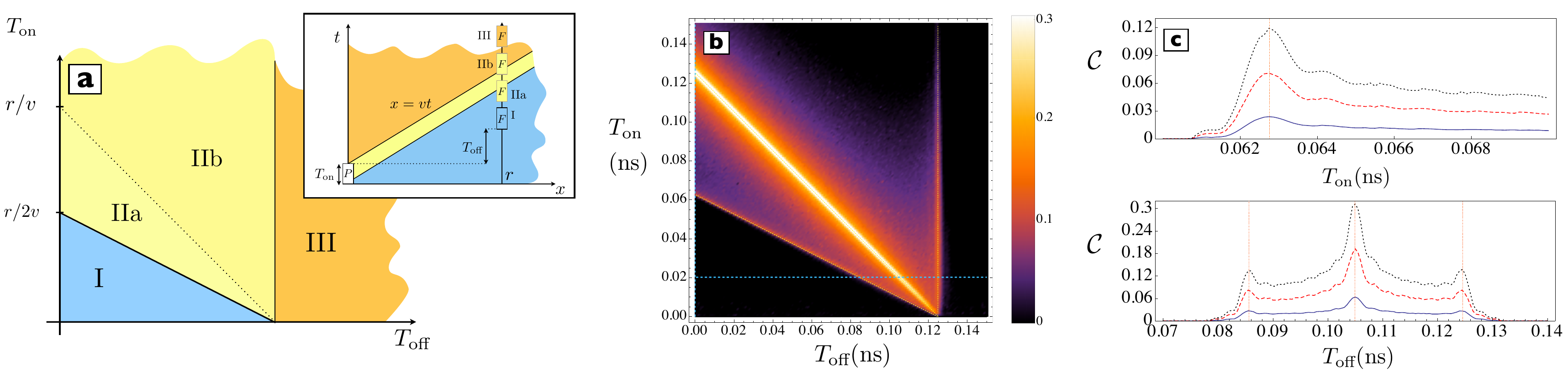}
  \caption{a) Diagram of the different spacetime regions.  b) Concurrence vs. $T_\text{on}$ and $T_\text{off}$ for  $g=g_P=g_F$, $\Omega=\Omega_P=\Omega_F=2\pi \times1\,\text{GHz}$, $g/\Omega=0.19$, $r/\lambda=0.125$. Significant entanglement is generated at both sides of the lines which discriminate between regions.  c) Concurrence vs. $T_\text{on}$ with $T_\text{off}$ fixed and viceversa along the blue lines shown in b. The peaks match the position of the region edges. Entanglement is generated in region I and III  for three different values of the coupling strength $g/\Omega=0.09$ (blue, solid), $0.15$ (red,dashed) and $0.19$ (black, dotted) and $T_\text{off}=0$. $T_\text{off}$ with $T_\text{on}=0.02\,\text{ns}$. $\Omega$ and $r$ are the same as in b. The generated entanglement displays a remarkable symmetry for regions I and III.}
  \label{fig:results}
\end{figure*}

Three different regions emerge from the  parameters above (see Fig.\ref{fig:results}a).  If $T_\text{off}<r/v$, we discriminate between two possibilities. First, if $2T_\text{on}+T_\text{off} <r/v$ (region I), there cannot be  real photon exchange, but vacuum correlations - or virtual photon exchange - are allowed at any time. If $2T_\text{on}+T_\text{off} >r/v$ (region II), $F$ may start to absorb radiation emitted by $P$ in the past somewhen in the future after an interval with no possible absorption (if $T_\text{on}+T_\text{off}<r/v$, region IIa) or start to absorb radiation at $t=t_1$ and stop to receive radiation somewhen in the future while the interaction is still on ($T_\text{on}+T_\text{off}>r/v$, region IIb). Finally, if $T_\text{off}>r/v$ (region III) $F$  cannot absorb radiation at all, as in region I. The difference between these two regions is that the qubits are spacelike separated in region I and timelike separated in region III. Only in regions I and III  we are dealing with a pure effect of transference of the past-future quantum correlations contained in the vacuum. In region II these correlations may be assisted by  a certain probability of photon exchange during a given time interval.

\begin{figure}[t]
 \vspace{0cm} 
\includegraphics[width=0.3\textwidth]{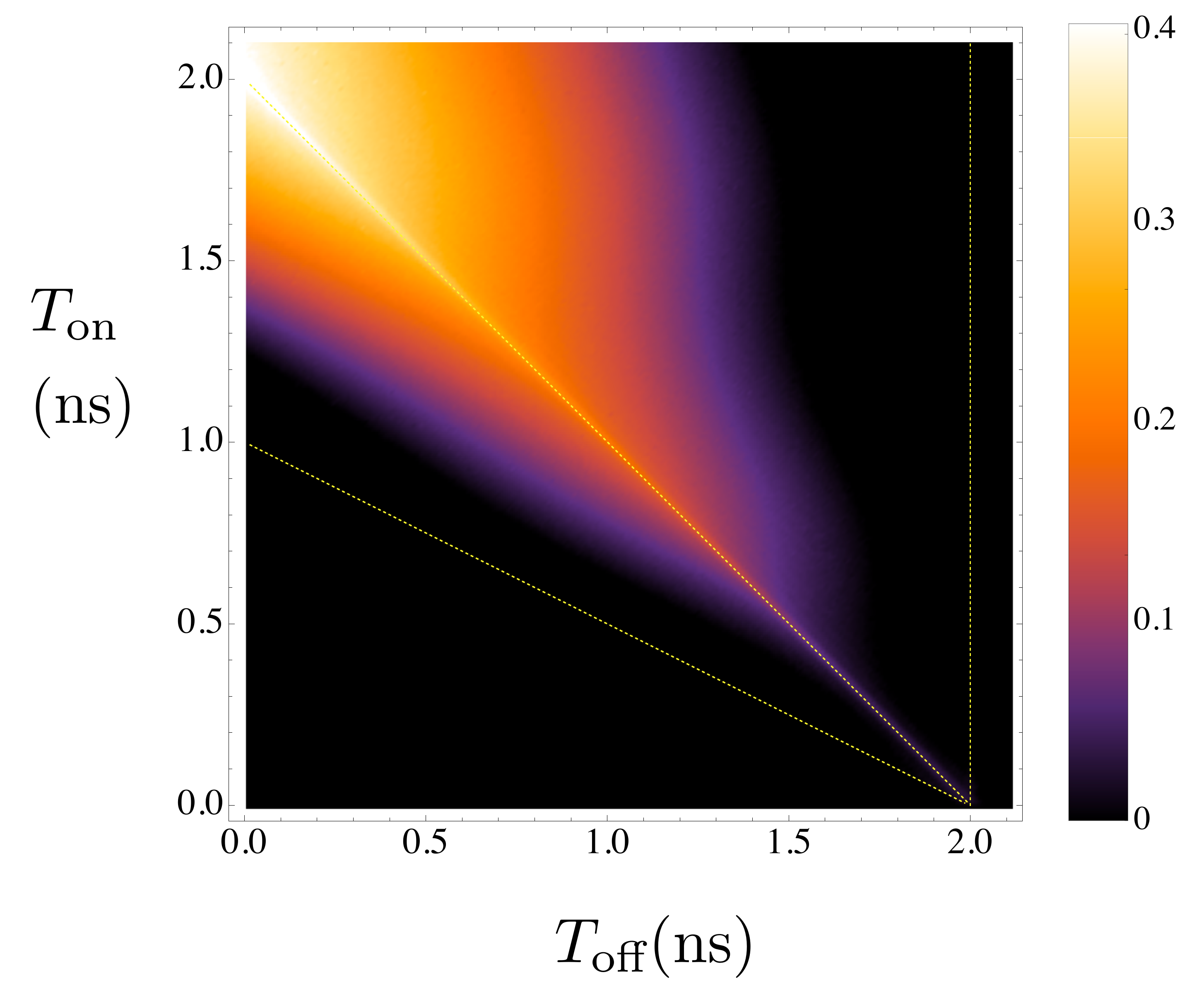}
\caption{ Same as in Fig.\ref{fig:results}b except for $r/\lambda= 2$ and $g/\Omega=0.09$. Entanglement is restricted to region II for long distances.}
\label{fig:results2}
\vspace{-0.5 cm}
\end{figure}

In Fig.~\ref{fig:results}b-c we show numerical results for the behavior of the concurrence as a function of $T_\text{on}$ and $T_\text{off}$, for coupling strengths $g_J/\Omega_J\simeq0.1$, such as in cutting-edge experiments of ultrastrong coupling in circuit QED \cite{forn-diaz10,niemczyk10} and accessible values of the qubit's gaps and distance. We note that qubit-qubit entanglement is  sizable in region II. However, the existence or not of entanglement in regions I and III depends much on the distance $r$, as expected. Fig.~\ref{fig:results}b-c show a certain amount of entanglement in regions I and III, entailing a pure transference of vacuum correlations. Remarkably, Fig. ~\ref{fig:results}c displays  an interesting symmetry between regions I and III: for a given interaction time the entanglement that can be generated only by transference of vacuum correlations is the same regardless whether the qubits are spacelike or timelike separated. This kind of entanglement vanishes as the distance grows (see Fig. \ref{fig:results2}).  In general, entanglement is concentrated around $\Omega_J\,T_\text{on}\simeq\Omega_J\,T_\text{off}\simeq1$ and $\xi_\text{on}\simeq\xi_\text{off}\simeq1$. Thus, for qubit distances of the order of $\lambda$ as in Fig. ~\ref{fig:results2}, entanglement shows up in the $\text{ns}$ regime, but drifts towards shorter times as the distance diminishes, as can be seen in Fig.~\ref{fig:results}b. 

From the experimental viewpoint, our protocol is equally interesting -and probably more amenable-  if the qubits are in region  II, although the origin of entanglement generation may seem at first glance less theoretically tantalizing. Notice however that even if photon exchange is allowed, our scheme does not include a projective measurement of the field state but a trace over all the field degrees of freedom instead. Then the generation of entanglement immediately after the light-cone crossing is not trivial. For instance, in the standard 3D matter-radiation Hamiltonian the atoms would only get classical correlations until much longer times \cite{conjuan}. Indeed the relationship of the light cone with entanglement without measurements is a peculiarity of circuit QED in the ultrastrong coupling regime, together with the very high degree of entanglement that can be achieved. Thus, even in region II, what we are introducing here is a novel way of entanglement generation, remarkably different from the standard ones, including quantum buses in superconducting cavities \cite{quantumbus}.

We note that in our scheme concurrence is 0 for $r=0$. This could seem at variance with the results in \cite{olsonralphpfextraction} where extraction of vacuum correlations to pair of timelike separated qubits in the same space point is reported. But notice that in \cite{olsonralphpfextraction} a tailor-made time-dependence for the qubit gap ($\propto 1/t$) is introduced, while in our scheme the gap is constant and we just switch on and off the interaction. As a matter of fact, the proposal in \cite{olsonralphpfextraction} exploits a formal analogy \cite{olsonralphfp}- only fulfilled for massless fields - between the past and the future light cones and the left-right Rindler wedges. However, one must be very careful about the extent to what this analogy is valid: while it is possible to think of the vacuum state as entangled in the modes observed by causally disconnected observers in the spacetime left-right wedges  \cite{Bruschi2010}, it is not clear whether this way of thinking can be transported to the past-future light cones \cite{comentario}. This was the reason of the singular -and arguably difficult to  implement experimentally- energy gap in \cite{olsonralphpfextraction}. However, we have shown that if the qubits are separated by a given distance $r$ and the interaction can be switched on and off fast enough to have finite interaction times, past-future entanglement can be generated between qubits with constant energy gaps.

\textit{Circuit QED realization.--} 
We will thus focus on the following setting, aiming to test the results shown in Fig.~\ref{fig:results}. As mentioned in the introduction, it consists of a circuit QED design where two superconducting qubits interact with the vacuum field in such a way that the interaction is on  during a finite time  \cite{peropadre10} and not at the same time for each qubit (see Fig.~\ref{fig:imp}a). After that, entanglement  can be quantified  with quantum state tomography \cite{tomographyscq}. 
This  qualitative scheme is based on switchable ultrastrong interactions that can be engineered using the design depicted in Fig.~[\ref{fig:imp}b]. A superconducting flux
qubit -upper three-junction loop-, is galvanically coupled to a
 quantum field $\Delta\psi$ -transmission line- by means of two
additional loops. These extra loops are essential since they will allow us to decouple the qubit from the field in an extremely fast way. 
We assume the Josephson energy $E_{J_{i}}$ of each junction to be much greater than its
charging energy $E_{C_{i}}$. Thus  the three-loop Hamiltonian
 can be reduced to a sum of the inductive
energies, $H_J=-\sum_{j=1}^5 E_{J_{j}}\cos\varphi_j$, where $\varphi_i$ is the superconducting phase of the i-th junction.

We can simplify the expression of $H_J$ due to the flux quantization around
each closed loop, that imposes:
${\modtwosum_j} \varphi_j=2\pi f_i,\quad(i=1,2,3),$ where the
magnetic frustration parameters $f_i=\Phi_{i}/\Phi_0$ depend on the
external magnetic fluxes $\Phi_i$.
Assuming the standard flux qubit configuration,
$E_{J1}=E_{J2}=\alpha E_{J3}$, together with
$E_{J4}=E_{J5}=\alpha_4E_{J1}$, the Hamiltonian is:
\begin{eqnarray}
H_J=&-E_J&\!\!\!\left [\cos(\varphi_1)+\cos(\varphi_2)+\alpha \cos(2\pi f_1+\varphi_1+\varphi_2) \right ]\nonumber\\
&-&\alpha_{\textrm{eff}}E_J\cos(2\pi f_{\textrm{eff}} \!-\!\Delta\psi+\varphi_1+\varphi_2).
\label{H}
\end{eqnarray}
where $\alpha_{\textrm{eff}}=2\alpha_4\cos(\pi f_3)$, and
$f_{\textrm{eff}}\!=\!f_1\!-\!f_2\!+\!f_3/2$ is an effective magnetic
frustration.  After a  proper diagonalization the first line of (\ref{H}) can be
identified with the flux qubit Hamiltonian,
whereas the second line represents the qubit-field interaction. The
shape of this interaction depends on 
 $f_1$ and $f_2$ and the interaction strength given by $\alpha_{\textrm{eff}}$, can
be adjusted through $f_3$. 
 A numerical evaluation of $H_J$ for $(f_1,f_2,f_3)=(0.5,0.75,1)$, yields the following effective
Hamiltonian  in the qubit basis:
$H=\int\!\!\! dk\omega_ka^\dagger a+\hbar\frac{\Omega}{2}\sigma_z+\alpha_4E_J\sigma_x\Delta\psi$, where $\Delta\psi$ is given by  (\ref{field}), and we have included the free Hamiltonian of the field. On the other hand, if we vary the SQUID magnetic flux up to $f_3\!\!=0.5$
the interaction is switched off, and the Hamiltonian is $H=\int\!\! dk\omega_ka^\dagger a+\hbar\frac{\Omega}{2}\sigma_z$. Therefore, with a fast change of $f_3$ the model given by  Eqs. (\ref{c})-(\ref{d}) can be realized in the laboratory. Current technology with Al qubits \cite{wilson} allows us to vary $f_3$ in times of less than 0.1 ns and this value can be further improved while remaining below the plasma frequency of the junctions.

\textit{Discussion.--} The extraction of past-future entanglement from the field to a pair of qubits could be used to implement a device which teleports a quantum state in time - as first suggested in \cite{olsonralphpfextraction}.  In other words, we could use the field in the  transmission line for building up a novel kind of quantum memory.  To achieve this goal, an observer - say, Paula -  in possession of $P$ and another qubit $P'$ that she wants to teleport,  carries out measurements on her qubits once the interaction is off at $-t_1$.  After $t_2$, an observer - say, Frank - would use the results of Paula's measurements stored as classical information and manipulate $F$ , in order to transfer the state of $P'$ to $F$. The fidelity will be a function of the amount of quantum correlations  between $P$ and $F$. Note that during $T_\text{off}$ the information of the state of $P'$ is  codified in the field, regardless whatever happened to $P$ after its interaction and measurement. The information is recovered and embodied in $F$ after its $T_\text{on}$ and the use of the stored classical bits. 

The experimental realization of quantum teleportation has already been achieved in cQED \cite{teleportationcircuits} and teleportation with mixed states is oncisdered in \cite{Sougatolet,TelepMixed}.  As shown in figure \ref{fig:results2}, entanglement is strong enough to consider high-fidelity teleportation for $T_\text{off}$ of nanoseconds.  This interval might in principle be even similar to the coherence times of the qubit and the scheme might be used as a quantum memory - provided that the coherence of the field is long enough. In our setting, that time-lapse grows with the qubit spatial separation and  the inverse of the qubit gap.

\textit{Conclusions.--}  We have proposed a circuit QED setup in which past-future  correlations can be transferred from a quantum field to a pair of qubits P and F, which only interact with the field in the past or the future respectively. We discuss  the possible technological uses of that entanglement extraction and the potential of our scheme to work as a quantum memory.

\textit{Acknowledgements.--} The authors would like to acknowledge J. J. Garc{\'\i}a-Ripoll, T. C. Ralph, A. Dragan,  I. Fuentes,  G. Adesso, J. Louko, N. Friis, C. M. Wilson, P. Delsing, P. Forn-D{\'\i}az, and J. Le{\'o}n for useful comments. This work was supported by Spanish MICINN Projects FIS2011-29287 and FIS2009-10061 and CAM research consortium QUITEMAD S2009-ESP-1594. B. Peropadre  and M. del Rey were supported by a CSIC JAE-PREDOC grant. M.del Rey was also supported by Ayuntamiento de Madrid.


\begin{thebibliography}{28}
\bibitem[{com({\natexlab{a}})}]{comment}
\bibinfo{howpublished}{Although in \cite{olsonralphfp} these correlations are
  referred to as entanglement, such a wording might not be appropriate since
  quantum states in past and future belong to Hilbert spaces defined in
  different simultaneity planes. Nevertheless, quantum correlations of this
  sort can in principle be extracted to physical systems and exploited as a
  resource for quantum information tasks as in the case of typical
  entanglement.}
\bibitem[{com({\natexlab{b}})}]{comment2}
\bibinfo{howpublished}{Note that we will use the term ``past-future''
  correlations to refer to a different and more general notion than ``time-like
  entanglement'' or ``entanglement between the past and the future light
  cones''.}
\bibitem[{com({\natexlab{c}})}]{comentario}
\bibinfo{howpublished}{To have a mirror interpretation for a plane-wave-like
  basis like that provided in \cite{olsonralphfp} we would need observers
  undergoing forbidden trajectories in the spacetime -horizontal hyperbolas in
  past and future light cones.}
\end{thebibliography}
\end{document}